\documentclass[dvips]{article}
\usepackage{supertabular,lscape,epsfig}
\usepackage{amssymb}
\usepackage{amsmath}
\usepackage{rotating}
\DeclareSymbolFont{ppa}{OT1}{ppl}{m}{it}
\DeclareMathSymbol{\vv}{\mathalpha}{ppa}{'166}

\begin{document}

\newcommand{\dd}{\,{\rm d}}
\newcommand{\ie}{{\it i.e.},\,}
\newcommand{\etal}{{\it et al.\ }}
\newcommand{\eg}{{\it e.g.},\,}
\newcommand{\cf}{{\it cf.\ }}
\newcommand{\vs}{{\it vs.\ }}
\newcommand{\zdot}{\makebox[0pt][l]{.}}
\newcommand{\up}[1]{\ifmmode^{\rm #1}\else$^{\rm #1}$\fi}
\newcommand{\dn}[1]{\ifmmode_{\rm #1}\else$_{\rm #1}$\fi}
\newcommand{\upd}{\up{d}}
\newcommand{\uph}{\up{h}}
\newcommand{\upm}{\up{m}}
\newcommand{\ups}{\up{s}}
\newcommand{\arcd}{\ifmmode^{\circ}\else$^{\circ}$\fi}
\newcommand{\arcm}{\ifmmode{'}\else$'$\fi}
\newcommand{\arcs}{\ifmmode{''}\else$''$\fi}
\newcommand{\MS}{{\rm M}\ifmmode_{\odot}\else$_{\odot}$\fi}
\newcommand{\RS}{{\rm R}\ifmmode_{\odot}\else$_{\odot}$\fi}
\newcommand{\LS}{{\rm L}\ifmmode_{\odot}\else$_{\odot}$\fi}

\newcommand{\Abstract}[2]{{\footnotesize\begin{center}ABSTRACT\end{center}
\vspace{1mm}\par#1\par
\noindent
{~}{\it #2}}}

\newcommand{\TabCap}[2]{\begin{center}\parbox[t]{#1}{\begin{center}
  \small {\spaceskip 2pt plus 1pt minus 1pt T a b l e}
  \refstepcounter{table}\thetable \\[2mm]
  \footnotesize #2 \end{center}}\end{center}}

\newcommand{\TableSep}[2]{\begin{table}[p]\vspace{#1}
\TabCap{#2}\end{table}}

\newcommand{\FigCap}[1]{\footnotesize\par\noindent Fig.\  %
  \refstepcounter{figure}\thefigure. #1\par}

\newcommand{\TableFont}{\footnotesize}
\newcommand{\TableFontIt}{\ttit}
\newcommand{\SetTableFont}[1]{\renewcommand{\TableFont}{#1}}

\newcommand{\MakeTable}[4]{\begin{table}[htb]\TabCap{#2}{#3}
  \begin{center} \TableFont \begin{tabular}{#1} #4
  \end{tabular}\end{center}\end{table}}

\newcommand{\MakeTableSep}[4]{\begin{table}[p]\TabCap{#2}{#3}
  \begin{center} \TableFont \begin{tabular}{#1} #4
  \end{tabular}\end{center}\end{table}}

\newenvironment{references}%
{
\footnotesize \frenchspacing
\renewcommand{\thesection}{}
\renewcommand{\in}{{\rm in }}
\renewcommand{\AA}{Astron.\ Astrophys.}
\newcommand{\AAS}{Astron.~Astrophys.~Suppl.~Ser.}
\newcommand{\ApJ}{Astrophys.\ J.}
\newcommand{\ApJS}{Astrophys.\ J.~Suppl.~Ser.}
\newcommand{\ApJL}{Astrophys.\ J.~Letters}
\newcommand{\AJ}{Astron.\ J.}
\newcommand{\IBVS}{IBVS}
\newcommand{\PASP}{P.A.S.P.}
\newcommand{\Acta}{Acta Astron.}
\newcommand{\MNRAS}{MNRAS}
\renewcommand{\and}{{\rm and }}
\section{{\rm REFERENCES}}
\sloppy \hyphenpenalty10000
\begin{list}{}{\leftmargin1cm\listparindent-1cm
\itemindent\listparindent\parsep0pt\itemsep0pt}}%
{\end{list}\vspace{2mm}}

\def\TYLDA{~}
\newlength{\DW}
\settowidth{\DW}{0}
\newcommand{\dw}{\hspace{\DW}}

\newcommand{\refitem}[5]{\item[]{#1} #2%
\def\REFARG{#3}\ifx\REFARG\TYLDA\else, {\it#3}\fi
\def\REFARG{#4}\ifx\REFARG\TYLDA\else, {\bf#4}\fi
\def\REFARG{#5}\ifx\REFARG\TYLDA\else, {#5}\fi.}

\newcommand{\Section}[1]{\section{#1}}
\newcommand{\Subsection}[1]{\subsection{#1}}
\newcommand{\Acknow}[1]{\par\vspace{5mm}{\bf Acknowledgements.} #1}
\pagestyle{myheadings}

\newfont{\bb}{ptmbi8t at 12pt}
\newcommand{\xrule}{\rule{0pt}{2.5ex}}
\newcommand{\xxrule}{\rule[-1.8ex]{0pt}{4.5ex}}
\def\thefootnote{\fnsymbol{footnote}}
\begin{center}

{\Large\bf
Photometric survey of stellar clusters in the outer part of M33. II. Analysis 
of HST/ACS  images}
\vskip1cm
{
\large K. Zloczewski and J. Kaluzny}
\vskip3mm
{
        Nicolaus Copernicus Astronomical Center,\\
        ul. Bartycka 18, 00-716 Warsaw, Poland\\
        e-mail: (kzlocz,jka)@camk.edu.pl\\
}
\end{center}

\Abstract{We have used deep ACS/WFC images of M33 to check nature
of extended objects detected by the ground based survey of 
Zloczewski et al. (2008).
A total of 24 candidates turned out to be genuine compact 
stellar clusters. In addition we detected 91 new clusters.
Equatorial coordinates, integrated magnitudes and  angular sizes
are listed for all 115 objects. Fourty-two clusters have sufficiently red colors
to be candidates for old globulars.  
For four clusters we extracted resolved 
stellar photometry.  Object 33-3-018 located in the outer disk of M33 
turned out to be a young cluster with an age estimated at 200-350 Myr.
Cluster ZK-90 has an age of 3-5 Gyr.
The remaining two clusters  have intermediate ages 
ranging from one to a few Gyr.}  
{catalogs-- galaxies: individual (M33) -- 
galaxies: star clusters}

\section{Introduction}

Several surveys have been made to catalog stellar clusters in M33.
In the last decade most of the clusters were found in the central part of
the galaxy using imaging instruments on board of the  Hubble Space Telescope
(Chandar et al. 1999, 2001, Bedin et al. 2005, Park and Lee 2007).
Wide-field searches for M33 clusters were also performed using
ground-based observations (Hintler 1960; Melnick and D'Odorico 1978;
Christian and Schommer 1982; Mochejska et al. 1998; Zloczewski et al.
2008, hereafter ZKH). The number of confirmed and candidate clusters 
is growing. Their
on-line catalog is provided by Sarajedini and Mancone 
(2007, hereafter SM). These authors point out that the available sample
is spatially incomplete. Indeed, only the central region of the galaxy
with a size of about 15 $\times$ 15 arcmin$^2$ was surveyed systematically
based on the HST WFPC2 images (see Fig. 1 of Park and Lee 2007).
It is also worth to note  that the faintest known M33 globular cluster 
candidates have $V \approx 21$  mag which corresponds 
to $M_V \approx -4$ (SM; ZKH). 
This can be compared with $M_V \approx -1$ observed for the faintest
of the known globulars in the Milky Way (Koposov et al. 2007). 
One may expect that several faint globular clusters from M33 still 
await  detection. 

During the last two years several HST/ACS images covering various locations
in the M33 have become publicly available. This prompted us to examine
in detail some of candidate clusters included in our catalog of 
extended objects from the outer part of the galaxy (ZKH). 
Results of this examination are presented in Section 2. As it turned out,
the analyzed images contain a few dozen of newly detected 
stellar clusters as well as several previously unconfirmed cluster
candidates. Their catalog as well as  integrated photometry are presented 
in Section 3. Section 4 is devoted to presentation and brief 
discussion of the 
color-magnitude diagrams (CMD) for 4 selected clusters. Summary 
and conclusions are given in Section 5.     

\section{Examination of ZKH cluster candidates}.

In our recent paper (ZKH) we presented a large sample of new candidate
stellar clusters from the outer part of M33. An area of $0.75$~deg$^2$  
was surveyed based on deep $g'r'i'$ images collected with the 
CFHT telescope. The images had a sub-arcsecond seeing and a 
scale of 0.185 arcsec/pixel.  Our catalog included 122 new likely 
compact stellar clusters, 3462 unclassified objects and 1155 galaxies. 
In addition,
we recovered 41 clusters listed as "confirmed" in the SM catalog.
Unclassified objects were selected in such a way that they
could be considered candidates for unresolved stellar clusters.  
 
Using HLA\footnote{http://hla.stsci.edu} archive we have identified and
imported images of seven M33 fields observed with the  ACS/WFC.
These particular fields overlap (6 totally and 1 partially) 
with the region surveyed in ZKH. All images were obtained in the 
WFC mode with the 0.05 arcsec/pixel scale. The used filters 
include F475W, F606W and F814W. The frames were collected
within programs
\#10190 (PI  Garnett), \#9480 (PI Rhodes) and \#9837 (PI Ferguson).
We note that these programs yielded also data for 7 fields located in the central part  
of M33. These  fields were not covered by ZKH survey (one of ACS fields
covering outskirts of M33 is also located out of ZKH survey area).

The examined ACS fields included 84 candidate clusters (unclassified
objects) and 11 likely clusters from ZKH. Out of  84 candidate clusters 64 
turned out to be  background galaxies and 4 still cannot
be classified with confidence. The remaining 16 candidates turned out to be 
genuine stellar clusters. As far as 11 likely clusters from ZKH
are considered, we confirmed the classification for 8 of them. 
Among the remaining three objects two 
turned out to be  galaxies and the nature of one remains unclear.
In Table 1 we provide new classifications for all 95 extended 
ZKH sources identified on the ACS images.  
Object name from the ZKH catalog is listed in the first column of Table 1.
Second column gives  the name of the ACS/WFC image on which a given object 
can be seen. Object types assigned in columns 3 and 4 follow the convention
adopted in  the ZKH catalog.

\begin{table}[h!]
\begin{center} \begin{tabular}{| c | c | c | r || c | c | c | r |}
\hline
ZHK	& name of  & ZHK  & HST  & ZHK      & name of  & ZHK  & HST\\
ID	& HST set  & type & type & ID       & HST set  & type & type\\
\hline
19-1-037 & j8g803031 & 0 & $-$1   & 25-1-010 & j90o37011 & 0 & $-$1\\
19-1-045 & j8g803031 & 0 & $-$1   & 25-1-011 & j90o37011 & 0 & $-$1\\
19-1-047 & j8g803031 & 0 & $-$1   & 25-1-013 & j90o37011 & 0 & $-$1\\
19-1-050 & j8g803031 & 0 & $-$1   & 25-1-015 & j90o37011 & 0 & $-$1\\
19-1-056 & j8g803031 & 0 & $-$1   & 32-5-017 & j90o47011 & 0 & $-$1\\
19-1-062 & j8g803031 & 0 & $-$1   & 32-5-018 & j90o47011 & 0 & $-$1\\
19-1-067 & j8g803031 & 0 & $-$1   & 32-5-021 & j90o47011 & 0 & $-$1\\
19-1-068 & j8g803031 & 0 & $-$1   & 32-5-022 & j90o47011 & 0 & $-$1\\
20-1-001 & j8g802031 & 0 & $-$1   & 32-5-024 & j90o47011 & 0 & 1\\
20-1-006 & j8g802031 & 0 & $-$1   & 33-2-001 & j90o47011 & 0 & 0\\
20-1-007 & j8g802031 & 0 & $-$1   & 33-2-003 & j90o47011 & 0 & 1\\
20-1-008 & j8g802031 & 0 & $-$1   & 33-2-005 & j90o47011 & 0 & $-$1\\
20-1-012 & j8g802031 & 0 & $-$1   & 33-2-006 & j90o47011 & 0 & 1\\
20-1-014 & j8g802031 & 0 & $-$1   & 33-2-007 & j90o47011 & 0 & $-$1\\
20-1-016 & j8g802031 & 0 & $-$1   & 33-2-010 & j90o47011 & 0 & 1\\
20-1-017 & j8g802031 & 0 & $-$1   & 33-3-003 & j90o47zbq & 0 & $-$1\\
20-1-018 & j8g802031 & 0 & $-$1   & 33-3-009 & j90o31011 & 1 & $-$1\\
20-1-020 & j8g802031 & 0 & $-$1   & 33-3-013 & j90o31011 & 0 & $-$1\\
20-1-021 & j8g802031 & 0 & $-$1   & 33-3-019 & j90o31011 & 1 & 0\\
20-1-022 & j8g802031 & 0 & $-$1   & 33-3-020 & j90o31011 & 0 & 1\\
20-1-023 & j8g802031 & 0 & $-$1   & 33-3-021 & j90o31011 & 1 & 1\\
20-1-024 & j8g802031 & 0 & $-$1   & 33-4-018 & j90o41011 & 1 & 1\\
20-1-025 & j8g802031 & 0 & $-$1   & 33-5-011 & j90o41011 & 0 & 1\\
20-1-026 & j8g801031 & 0 & $-$1   & 33-5-013 & j90o41011 & 0 & 1\\
20-1-027 & j8g802031 & 0 & $-$1   & 33-5-014 & j90o41011 & 0 & 1\\
20-1-031 & j8g801031 & 0 & $-$1   & 33-5-015 & j90o41011 & 0 & 1\\
20-4-002 & j8g801031 & 0 & $-$1   & 33-5-019 & j90o41011 & 0 & 1\\
20-4-003 & j8g801031 & 0 & $-$1   & 33-5-020 & j90o41011 & 0 & 0\\
20-4-007 & j8g801031 & 0 & $-$1   & 33-5-021 & j90o41011 & 0 & $-$1\\
20-4-008 & j8g801031 & 0 & $-$1   & 33-5-022 & j90o41011 & 0 & 1\\
20-4-011 & j8g801031 & 0 & $-$1   & 33-5-023 & j90o41011 & 0 & $-$1\\
20-4-015 & j8g801031 & 0 & $-$1   & 33-6-002 & j90o31011 & 0 & $-$1\\
20-4-017 & j8g801031 & 0 & $-$1   & 33-6-003 & j90o31011 & 0 & $-$1\\
20-4-019 & j8g801031 & 0 & $-$1   & 33-6-005 & j90o31011 & 0 & $-$1\\
20-4-020 & j8g801031 & 0 & $-$1   & 33-6-007 & j90o48a9q & 0 & 0\\
20-4-022 & j8g801031 & 0 & $-$1   & 33-6-008 & j90o31011 & 0 & 1\\
20-4-024 & j8g801031 & 0 & $-$1   & 33-6-009 & j90o31011 & 1 & 1\\
20-4-027 & j8g801031 & 0 & $-$1   & 33-6-010 & j90o31011 & 0 & 1\\
20-4-028 & j8g801031 & 0 & $-$1   & 33-6-012 & j90o31011 & 1 & 1\\
20-5-021 & j8g801031 & 0 & $-$1   & 33-6-014 & j90o31011 & 1 & 1\\
25-1-001 & j90o37011 & 0 &    1   & 33-6-016 & j90o48030 & 1 & $-$1\\
25-1-002 & j90o37011 & 0 & $-$1   & 33-6-017 & j90o48030 & 1 & 1\\
25-1-003 & j90o37011 & 0 & 1      & 33-6-019 & j90o31050 & 1 & 1\\
25-1-004 & j90o37011 & 0 & $-$1   & 33-6-020 & j90o48a9q & 0 & 0\\
25-1-005 & j90o37011 & 0 & $-$1   & 34-1-002 & j90o41011 & 0 & $-$1\\
25-1-007 & j90o37011 & 0 & $-$1   & 34-1-004 & j90o41011 & 0 & $-$1\\
25-1-008 & j90o37011 & 0 & 1      & 34-2-001 & j90o41011 & 1 & 1\\
25-1-009 & j90o37011 & 0 & $-$1   & & & &\\
\hline
\end{tabular}
\caption{New classification of 95 objects from the ZKH catalog. Object
types listed in
columns 3/7 and 4/8 have the following meaning: -1 galaxy, 0 unclassified,
1 probable cluster. Objects with HST type set to 1 can be considered certain
clusters. The data are available in electronic form
at http://case.camk.edu.pl/results/index.html.
}
\end{center}
\end{table}

\Section{Newly identified clusters}

While checking the status of objects  listed by ZKH  we noted
that the examined images contain some previously uncatalogued 
stellar clusters. Using images of all 15 fields observed within the programs
listed in Sec. 2
we identified a total of 91 new stellar clusters. In addition, we were able
to confirm cluster nature of one object (SM 57 = ZKH 33-3-021) which is not
included on the  list of "high confidence clusters" in the SM catalog.
Positions and integrated photometry for these objects
is provided in Table 2. The first 24 entries correspond to clusters
from the ZKH list.
The photometry is on the VEGAMAG system defined in Sirianni et al. (2005).
The aperture magnitudes were extracted using Daophot code (Stetson 1987).
The same aperture radius was used for all
filters while extracting photometry for a given cluster.
In Table 2 cluster IDs are followed by their equatorial coordinates.
Column 4 provides the name of a data set on which a given object can be 
located, while columns 5-7 give VEGAMAG magnitudes for 3 considered bands. 
The aperture radius used for extraction of integrated photometry is 
listed in column 8. Its value is equal to the 
estimated angular radius of a given cluster.
Last column gives projected distance from the center of M33
($RA=23.46212$~deg, $DEC=+30.66028$~deg, J2000.0). \\

\begin{table}[h!]
 \begin{tabular}{| r c c c | c c c | c  r |}
\hline
ID & ${\alpha}_{2000.0}$[$^\circ$] & ${\delta}_{2000.0}$[$^\circ$] & \small{HST dataset} & \small{F475W} & \small{F606W}
& \small{F814W} & \small{ap} [$''$] & d [$'$] \\
\hline
25-1-008 & 23.24984 & 30.45544 & j90o37011 & -- & 20.51 & 19.50 & 1.00 & 16.5\\
25-1-003 & 23.26001 & 30.44884 & j90o37011 & -- & 20.41 & 20.09 & 1.15 & 16.4\\
25-1-001 & 23.26241 & 30.44294 & j90o37011 & -- & 20.16 & 19.60 & 1.00 & 16.6\\
34-2-001 & 23.26375 & 30.26727 & j90o41011 & 20.70 & 20.27 & 19.69 & 2.15 & 25.7\\
33-4-018 & 23.27343 & 30.23981 & j90o41011 & 19.61 & 19.19 & 18.68 & 2.75 & 27.1\\
33-6-019 & 23.28187 & 30.32722 & j90o31050 & -- & -- & -- & -- & 22.0\\
33-6-017 & 23.28519 & 30.39167 & j90o48030 & -- & 20.03 & 19.48 & 1.50 & 18.5\\
33-5-022 & 23.28724 & 30.28100 & j90o41011 & 21.92 & 21.34 & 20.73 & 0.80 & 24.5\\
33-5-019 & 23.29325 & 30.26272 & j90o41011 & 21.86 & 21.16 & 20.37 & 1.05 & 25.4\\
33-6-014 & 23.30086 & 30.37688 & j90o31011 & 20.59 & 19.84 & 18.94 & 1.50 & 18.9\\
33-6-012 & 23.30160 & 30.36632 & j90o31011 & 20.22 & 19.57 & 18.80 & 2.00 & 19.5\\
33-6-010 & 23.30420 & 30.38527 & j90o31011 & 19.37 & 19.40 & 19.47 & 0.80 & 18.4\\
33-5-015 & 23.30471 & 30.26116 & j90o41011 & 21.86 & 21.45 & 21.06 & 1.20 & 25.3\\
33-5-014 & 23.30814 & 30.24217 & j90o41011 & 20.96 & 20.30 & 19.64 & 1.75 & 26.3\\
33-5-013 & 23.30848 & 30.25453 & j90o41011 & 20.84 & 20.50 & 20.04 & 1.50 & 25.6\\
33-6-009 & 23.31157 & 30.38852 & j90o31011 & 19.83 & 19.23 & 18.54 & 2.00 & 18.1\\
33-5-011 & 23.31179 & 30.24392 & j90o41011 & 20.72 & 20.40 & 20.09 & 1.65 & 26.2\\
33-6-008 & 23.31341 & 30.35390 & j90o31011 & 19.68 & 19.57 & 19.46 & 1.75 & 19.9\\
33-3-021 & 23.33005 & 30.38955 & j90o31011 & 19.01 & 18.23 & 17.58 & 2.00 & 17.6\\
33-3-020 & 23.33879 & 30.34219 & j90o31011 & 18.59 & 18.36 & 18.06 & 1.50 & 20.1\\
33-2-010 & 23.35720 & 30.30032 & j90o47011 & -- & 20.56 & 20.51 & 1.00 & 22.3\\
33-2-006 & 23.37443 & 30.30926 & j90o47011 & -- & 19.83 & 19.64 & 1.15 & 21.5\\
33-2-003 & 23.38699 & 30.26299 & j90o47011 & -- & 20.23 & 20.04 & 1.10 & 24.2\\
32-5-024 & 23.40152 & 30.25905 & j90o47011 & -- & 21.57 & 21.20 & 0.75 & 24.3\\
\hline
ZK-1 & 23.27445 & 30.48041 & j90o37011 & -- & 19.71 & 19.35 & 1.20 & 14.5\\
ZK-2 & 23.28861 & 30.43520 & j90o37011 & -- & 19.91 & 19.46 & 1.10 & 16.2\\ 
ZK-3 & 23.29808 & 30.26763 & j90o41011 & 21.37 & 21.22 & 21.01 & 1.25 & 25.0\\
ZK-4 & 23.29927 & 30.43011 & j90o37011 & -- & 20.48 & 20.01 & 1.00 & 16.2\\ 
ZK-5 & 23.30804 & 30.46668 & j90o37011 & -- & 19.41 & 18.82 & 1.50 & 14.1\\ 
ZK-6 & 23.32407 & 30.38805 & j90o31011 & 22.12 & 21.50 & 20.79 & 0.80 & 17.8\\
ZK-7 & 23.33642 & 30.34723 & j90o31011 & 21.77 & 21.31 & 20.79 & 0.85 & 19.9\\
ZK-8 & 23.33880 & 30.51694 & j90o38030 & -- & 20.06 & 19.28 & 0.75 & 10.7\\ 
ZK-9 & 23.34221 & 30.64115 & j90o13011 & -- & 19.07 & 18.24 & 1.70 & 6.3\\ 
ZK-10 & 23.34440 & 30.63339 & j90o13011 & -- & 20.61 & 19.84 & 0.75 & 6.3\\
ZK-11 & 23.34484 & 30.63883 & j90o13011 & -- & 19.62 & 18.81 & 1.00 & 6.2\\ 
ZK-12 & 23.34813 & 30.49819 & j90o38030 & -- & 20.90 & 20.31 & 0.75 & 11.4\\
ZK-13 & 23.34936 & 30.66017 & j90o13011 & -- & 18.95 & 18.46 & 1.65 & 5.8\\ 
ZK-14 & 23.35041 & 30.52869 & j90o38030 & -- & 19.90 & 19.24 & 1.20 & 9.8\\ 
\hline
\end{tabular}
\caption{Properties of 115 stellar clusters from M33.}
\end{table}

\setcounter{table}{1}

\begin{table}[h!]
 \begin{tabular}{| r c c c | c c c | c  r |}
\hline
ID & ${\alpha}_{2000.0}$[$^\circ$] & ${\delta}_{2000.0}$[$^\circ$] & \small{HST dataset} & \small{F475W} & \small{F606W}
& \small{F814W} & \small{ap} [$''$] & d [$'$] \\
\hline
ZK-15 & 23.35252 & 30.63051 & j90o13011 & -- & 19.24 & 18.58 & 1.45 & 5.9\\
ZK-16 & 23.35474 & 30.53838 & j90o38030 & -- & 20.47 & 19.72 & 1.10 & 9.2\\
ZK-17 & 23.35552 & 30.39420 & j90o31011 & 21.38 & 20.96 & 20.52 & 0.90 & 16.9\\   
ZK-18 & 23.35722 & 30.52218 & j90o38030 & -- & 20.47 & 19.98 & 0.85 & 9.9\\
ZK-19 & 23.35805 & 30.52263 & j90o38030 & -- & 20.29 & 19.36 & 1.00 & 9.9\\
ZK-20 & 23.36008 & 30.66655 & j90o13011 & -- & 19.62 & 18.90 & 0.75 & 5.3\\
ZK-21 & 23.36422 & 30.50099 & j90o38030 & -- & 19.57 & 18.70 & 1.00 & 10.8\\
ZK-22 & 23.36608 & 30.64695 & j90o13011 & -- & 19.38 & 18.84 & 1.20 & 5.0\\   
ZK-23 & 23.37824 & 30.48730 & j90o22011 & -- & 18.57 & 18.08 & 0.90 & 11.2\\
ZK-24 & 23.38097 & 30.47834 & j90o22011 & -- & 18.55 & 17.87 & 1.40 & 11.7\\
ZK-25 & 23.38496 & 30.29347 & j90o47011 & -- & 21.22 & 20.60 & 0.95 & 22.4\\  
ZK-26 & 23.38906 & 30.49606 & j90o22011 & -- & 19.69 & 19.54 & 0.90 & 10.6\\  
ZK-27 & 23.39010 & 30.46933 & j90o22011 & -- & 18.35 & 18.11 & 1.30 & 12.0\\
ZK-28 & 23.39034 & 30.53376 & j90o38030 & -- & 18.64 & 18.58 & 1.50 & 8.4\\
ZK-29 & 23.39116 & 30.66503 & j90o13011 & -- & 18.85 & 18.36 & 1.25 & 3.7\\
ZK-30 & 23.39269 & 30.48331 & j90o22011 & -- & 20.03 & 19.02 & 1.35 & 11.2\\
ZK-31 & 23.39943 & 30.46234 & j90o22011 & -- & 20.68 & 20.45 & 0.70 & 12.3\\
ZK-32 & 23.40098 & 30.46576 & j90o22011 & -- & 19.40 & 18.76 & 1.40 & 12.1\\
ZK-33 & 23.40611 & 30.44973 & j90o22011 & -- & 20.34 & 19.99 & 0.75 & 13.0\\
ZK-34 & 23.40670 & 30.47824 & j90o22011 & -- & 20.88 & 20.36 & 1.00 & 11.3\\
ZK-35 & 23.40805 & 30.31439 & j90o47011 & -- & 20.14 & 19.65 & 1.25 & 20.9\\
ZK-36 & 23.41447 & 30.46871 & j90o22011 & -- & 20.05 & 19.01 & 1.00 & 11.8\\
ZK-37 & 23.41731 & 30.62925 & j90o28050 & -- & 18.31 & 17.74 & 1.25 & 3.0\\
ZK-38 & 23.41835 & 30.59441 & j90o11011 & -- & -- & -- & -- & 4.6\\
ZK-39 & 23.41958 & 30.30716 & j90o47011 & -- & 20.84 & 20.62 & 1.25 & 21.3\\
ZK-40 & 23.42263 & 30.52041 & j90o22011 & -- & 18.45 & 18.28 & 1.35 & 8.6\\
ZK-41 & 23.42401 & 30.49234 & j90o22011 & -- & 19.15 & 18.67 & 0.75 & 10.3\\
ZK-42 & 23.42423 & 30.49280 & j90o22011 & -- & 19.87 & 19.75 & 0.75 & 10.2\\
ZK-43 & 23.42435 & 30.60058 & j90o11011 & -- & 18.76 & 18.21 & 1.05 & 4.1\\
ZK-44 & 23.42504 & 30.63887 & j90o28050 & -- & 18.65 & 17.95 & 1.25 & 2.3\\
ZK-45 & 23.42753 & 30.58304 & j90o11011 & -- & 19.91 & 19.02 & 1.00 & 5.0\\
ZK-46 & 23.42772 & 30.63930 & j90o28050 & -- & 18.74 & 18.20 & 1.00 & 2.2\\ 
ZK-47 & 23.42870 & 30.46291 & j90o22011 & -- & 20.12 & 19.73 & 1.00 & 12.0\\
ZK-48 & 23.42885 & 30.64173 & j90o28050 & -- & 19.04 & 18.58 & 0.80 & 2.0\\ 
ZK-49 & 23.43114 & 30.46896 & j90o22011 & -- & 19.59 & 18.89 & 1.00 & 11.6\\
ZK-50 & 23.43136 & 30.46618 & j90o22011 & -- & -- & -- & -- & 11.8\\
ZK-51 & 23.43299 & 30.60363 & j90o11011 & -- & 19.69 & 19.13 & 0.75 & 3.7\\ 
ZK-52 & 23.43400 & 30.59037 & j90o11011 & -- & 19.16 & 18.48 & 1.20 & 4.4\\   
ZK-53 & 23.43584 & 30.62600 & j90o28050 & -- & 19.20 & 18.69 & 1.00 & 2.5\\ 
\hline
\end{tabular}
\caption{Continued.}
\end{table}

\setcounter{table}{1}

\begin{table}[h!]
 \begin{tabular}{| r c c c | c c c | c  r |}
\hline
ID & ${\alpha}_{2000.0}$[$^\circ$] & ${\delta}_{2000.0}$[$^\circ$] & \small{HST dataset} & \small{F475W} & \small{F606W}
& \small{F814W} & \small{ap} [$''$] & d [$'$] \\
\hline
ZK-54 & 23.43603 & 30.60989 & j90o11011 & -- & 18.38 & 18.22 & 0.90 & 3.3\\ 
ZK-55 & 23.43675 & 30.57749 & j90o11011 & -- & 18.88 & 18.65 & 1.05 & 5.1\\   
ZK-56 & 23.44052 & 30.58011 & j90o11011 & -- & 18.44 & -- & 0.70 & 4.9\\      
ZK-57 & 23.44390 & 30.57840 & j90o11011 & -- & 19.04 & 18.10 & 1.70 & 5.0\\ 
ZK-58 & 23.44463 & 30.60324 & j90o11011 & -- & 19.87 & 19.27 & 1.05 & 3.5\\ 
ZK-59 & 23.44480 & 30.59973 & j90o11011 & -- & 19.66 & 18.83 & 1.00 & 3.7\\ 
ZK-60 & 23.45235 & 30.61812 & j90o11011 & -- & 18.32 & 17.52 & 1.15 & 2.6\\ 
ZK-61 & 23.45237 & 30.58981 & j90o11011 & -- & 19.42 & 18.83 & 1.15 & 4.3\\ 
ZK-62 & 23.45638 & 30.47844 & j90o22011 & -- & 19.90 & 19.44 & 1.00 & 10.9\\
ZK-63 & 23.46244 & 30.59012 & j90o11011 & -- & 20.09 & 19.41 & 0.75 & 4.2\\ 
ZK-64 & 23.47172 & 30.59045 & j90o11011 & -- & -- & -- & -- & 4.2\\
ZK-65 & 23.47211 & 30.52424 & j90o27011 & -- & 19.92 & 19.70 & 0.80 & 8.2\\ 
ZK-66 & 23.47246 & 30.58260 & j90o11011 & -- & 19.48 & 18.85 & 0.80 & 4.7\\
ZK-67 & 23.47381 & 30.53706 & j90o27011 & -- & 19.98 & 19.16 & 0.90 & 7.4\\
ZK-68 & 23.47563 & 30.60187 & j90o11011 & -- & 20.11 & 19.63 & 0.70 & 3.6\\
ZK-69 & 23.47662 & 30.53999 & j90o27011 & -- & 19.25 & 18.54 & 1.25 & 7.3\\
ZK-70 & 23.49103 & 30.53941 & j90o27011 & -- & 19.37 & -- & 1.10 & 7.4\\
ZK-71 & 23.49758 & 30.53338 & j90o27011 & -- & 19.36 & -- & 1.00 & 7.8\\    
ZK-72 & 23.49809 & 30.60674 & j90o11011 & -- & 19.67 & 19.47 & 0.55 & 3.7\\ 
ZK-73 & 23.49878 & 30.53337 & j90o27011 & -- & 19.77 & 19.04 & 1.00 & 7.8\\ 
ZK-74 & 23.50726 & 30.56839 & j90o27011 & -- & 19.23 & 18.50 & 1.50 & 6.0\\
ZK-75 & 23.50785 & 30.57299 & j90o27011 & -- & 18.81 & 17.86 & 1.70 & 5.7\\ 
ZK-76 & 23.50936 & 30.54385 & j90o27011 & -- & 19.05 & 18.13 & 1.15 & 7.4\\ 
ZK-77 & 23.51365 & 30.46550 & j90o14050 & -- & 17.81 & 17.40 & 2.35 & 12.0\\
ZK-78 & 23.51431 & 30.56153 & j90o27011 & -- & 19.71 & 19.40 & 0.95 & 6.5\\
ZK-79 & 23.51929 & 30.45594 & j90o14050 & -- & 19.65 & 19.06 & 2.15 & 12.6\\
ZK-80 & 23.52071 & 30.47216 & j90o14050 & -- & 20.50 & 19.74 & 1.00 & 11.7\\
ZK-81 & 23.52800 & 30.46783 & j90o14050 & -- & 19.47 & 18.91 & 1.30 & 12.0\\
ZK-82 & 23.52839 & 30.56815 & j90o27011 & -- & 19.84 & 19.03 & 1.00 & 6.5\\
ZK-83 & 23.52988 & 30.56281 & j90o27011 & -- & 19.41 & 18.76 & 1.50 & 6.8\\
ZK-84 & 23.54026 & 30.41803 & j90o14050 & -- & 20.02 & 19.75 & 1.25 & 15.1\\
ZK-85 & 23.54544 & 30.47564 & j90o14050 & -- & 19.77 & 19.50 & 1.40 & 11.9\\
ZK-86 & 23.55380 & 30.47947 & j90o14050 & -- & 18.13 & 17.65 & 1.50 & 11.8\\
ZK-87 & 23.55629 & 30.47884 & j90o14050 & -- & 19.81 & 19.15 & 1.60 & 11.9\\
ZK-88 & 23.57400 & 30.45241 & j90o14050 & -- & 20.82 & 20.44 & 0.85 & 13.7\\
ZK-89 & 23.57770 & 30.45526 & j90o14050 & -- & 20.05 & 19.34 & 1.15 & 13.7\\
ZK-90 & 23.75930 & 31.23930 & j8q802010 & -- & 20.66 & 19.84 & 2.40 & 38.0\\
ZK-91 & 23.76993 & 31.19940 & j8q802010 & -- & 21.89 & 21.64 & 1.00 & 36.0\\
\hline
\end{tabular}
\caption{Continued.}
\end{table}

Figure 1  shows color-magnitude diagram and 
color-color diagram for the clusters from Table 2. 
Two colors were available for 19 clusters only.
One may see that the measured colors span a
rather wide ranges. As for the magnitudes, the presented sample includes
some of the faintest stellar clusters identified so far in M33. 
We list a total of 36 objects with  $V>20${\footnote{$V$ magnitudes were estimated from F606W and F814W magnitudes using relations from Sirianni et al. 2005).} 
of which 9 have $21<V<22.0$. The recent catalog of Park \& Lee (2007)
includes just 8 clusters with $20<V<21.0$ and none with $V>21.0$.
Population of globular clusters from the Milky Way includes only old
objects with ages exceeding about 10 Gyr. In general 
they show rather red integrated colors. 
The blue edge of unreddened color distribution occurs at 
$B-V\approx 0.55$ and $V-I\approx 0.84$ (Park \& Lee 2007; 
Chandar et al. 2001). After correcting for the foreground extinction
of $E(B-V)=0.05$ (Schlegel et al. 1998) these colors transform
to $F475W-F606W=0.64 $ and $F606W-F814W=0.63 $. These limiting  values 
are marked with dotted lines in Fig. 1. One may see that our
sample  includes about 42 clusters from M33 with colors falling into 
the range observed for the Galactic globular clusters.

\section{Analysis of  selected  clusters}
Four clusters from Table 2 were chosen for a more detailed
analysis. We focused our attention on low surface density objects
located in the outer part of the galaxy. Their F814W 
filter images
are presented in Fig. 2. Stellar profile photometry was extracted 
from individual {\sc FLT} images using  
{\sc Dolphot} package (Dolphin, 2000). In each case reductions
were limited to the field of a size $800\times 800$~pixel$^{2}$ 
($40\times 40$~arcsec$^{2}$) field.
If possible, these
sub-frames were centered on the analyzed cluster. 
We followed the procedure 
described in {\sc Dolphot} manual. A predetermined list
of stellar positions was prepared by running {\sc daophot/allstar}
(Stetson, 1987) on deep drizzled image in the F814W band.
Pixel area map correction and bad pixels
maps were applied to {\sc  FLT} images.
Revised HST/ACS
zero-points\footnote{HST/ACS STScI webpage:
http://www.stsci.edu/hst/acs/analysis/zeropoints} were used to derive
calibrated photometry on the VEGAMAG system 
for F475W, F606W and F814W bands.
 
In the following analysis we adopt for the
foreground reddening  $E(B-V)=0.05$ (Schlegel 1998) for the M33 field.
The corresponding
extinction for ACS/WFC bands follows from relations given in 
Sirianni et al. (2005). As for the distance modulus, we adopt 
$(m-M)_{0}=24.64$ (Galleti et al. 2004). 

\Subsection{Cluster ZK-90}
With a projected angular distance from the galaxy center of $d=38.0'$,  
the cluster ZK-90 is located in the far outskirts of M33. The only
known cluster located at a larger galactocentric distance is    
M33-EC1 (Stonkute, 2008) with  $d=53'$.
ZK-90 is also the faintest of all cataloged M33 stellar clusters. 
Its apparent magnitudes, obtained using
calibrations from Sirianni et al. (2005) and photometry from Table 2 
are $V=20.90$ and $I=19.80$. 
The unreddened integrated magnitude 
and color of the cluster can be estimated at  
$V_{0}=20.74$ and $(V-I)_{0}=1.00$. This  implies  
an absolute magnitude  $M_{V}=-3.90$.\\
Figure 3 shows the surface density profile of ZK-90 derived
for stars with $F606W<27.5$. No correction for the incompleteness of 
photometry was done. The cluster radius can be estimated conservatively
at 2.2 arcsec, i.e.  $\sim$10 pc. Panel (a) of Fig. 4 shows CMD for 
the $40\times40$~arcsec$^{2}$ field including the 
cluster. The prominent red giant clump (RC) is visible at 
$F606W=25.05\pm 0.16$ and $F606W-F814W=0.74\pm 0.05$. The subgiant 
branch can be 
easily traced down to $F606W\approx 27.5$, some 2.5 mag below the red 
clump. 
Stars forming these features belong to old or intermediate
age populations (ages over 1 Gyrs). At the same time the brightest 
main sequence stars are observed at $F606W \approx 24.0$ which 
corresponds to an absolute magnitude $M_{\rm V}\approx -0.6$. 
Comparison  with available stellar models shows that
such bright and blue stars cannot be older than about 350 Myr. 
The CMD from Fig. 4a 
closely resembles CMD's for some other outer fields in M33 which
were analyzed in detail by Barker et al. (2007). 

The CMD for the region occupied by ZK-90 is shown if Fig. 4b (all stars are 
located at projected distance $d<2.5$~arcsec from the cluster center). 
In Fig. 4c we show the diagram for the "comparison" field which 
is located close to ZK-90 and has the same area as the cluster field.
Finally, Fig. 4d presents the cleaned CMD of the cluster
from which the field interlopers were statistically removed.
To obtain it, for each star
from panel (c) the nearest counterpart was located 
in panel (b) and then removed. The cleaned CMD 
of the cluster shows 10-11 RC stars and 3
candidates for bright red giants.
Subgiant branch
and upper main sequence are rather poorly defined. Three isochrones
plotted in Fig. 4d correspond to ages of 3, 4 and 5 Gyr (from left to right). 
They were extracted
from Dartmouth Stellar Evolution Database  (Dotter et al. 2008)
for ${\rm [Fe/H]}=-1.0$ and ${\rm [\alpha/Fe]}=0.2$. 
Isochrones for ${\rm [Fe/H]}=-0.5$ are too blue to match the 
observations. 
To transform model isochrones
to the observational plane we adopted $E(F606W-F814W)=0.05$ $A_{F606W}=0.14$
(they follow from $E(B-V)=0.05$ (Schlegel et al 1998)) 
and $(m-M)_{0}=24.64$ (Galleti et al. 2004). It may be concluded that the 
ZK-90  has an age  between 3 and 5 Gyrs and its metallicity is
close to ${\rm [Fe/H]}=-1.0$. 

A more detailed analysis of the 
cluster CMD is hampered by several difficulties. First of 
all,  photometry of main-sequence stars is  severely incomplete 
in the cluster field due to increased crowding. Moreover, comparisons
of panels $b$ and $c$ of Fig. 4 shows that field stars 
significantly contaminate the cluster region. We note that  ZK-90
contains about 10 red clump giants.  It is therefore richer
than well studied old open clusters from the solar vicinity like M67,
NGC~188, NGC~2243 
or Berkeley~39 (see CMD's in Carraro et al. (1994)).

\Subsection{Cluster 33-4-018}
The cluster 33-4-018 is located at a projected distance of 27.1 arcmin
from the center of M33. It is visible against the
faint outer spiral arm labeled as IV by Sandage  \& Humpherys (1980). 
It was included in the catalog of planetary nebula
candidates (Magrini et al. 2001; object ID=11)  based on the
presence of ${\rm H}\alpha $ and {\rm OIII} emission. 
On the composed color image HST\_10190\_46\_ACS\_WFC\_F814W\_F475W from the HST
Legacy Archive the object appears as a resolved stellar
cluster hosting several bright blue stars. 
Upper panels of Fig. 5 show
CMDs for the $40\times 40$~arcsec$^{2}$ field centered on the cluster.
Stars marked with large symbols are located at a distance
$r<2.0$~arcsec from the cluster center.
The bottom panel of Fig. 5 shows CMD for a nearby "comparison" field 
having the same area as the considered cluster region with 
$r<2.0$~arcsec. 

It is evident that the cluster region shows an excess of
bright stars with $F606W<24.5$ relatively to the comparison field.
These  stars are located on the upper main sequence as well as
among  bright red giants. The field population contains
some main sequence stars as bright as $F606W\approx 22.5$ or 
$M_{\rm V}\approx -2.3$. However, as can be seen in Figs. 5cd,
bright main sequence stars with $F606W<24.5$ or $M_{\rm V} <-0.3$ are generally 
rare in this part of M33.
One may also notice in Figs. 5ab that the completeness of cluster photometry 
diminishes rapidly for $F606W>24.5$. This is due to crowding effect, which
makes the detection of fainter stars in a region
filled with brighter objects practically impossible. 
Figure 6 presents CMD of the cluster                    
cleaned the same way as before from the field interlopers.
An apparent lack of stars with $F606W<26.5$ is entirely due to 
the incompleteness effect resulting in an "over-subtraction"
of field stars from the cluster CMD.\\
We have attempted to estimate the age of the cluster by fitting 
model isochrones from Dotter et al. (2008). 
Since there is no  spectroscopic information about the metallicity of 
33-4-018, we considered models with three values of 
[Fe/H]: 0.0, $-0.5$ and $-1.0$. The isochrones for ages of 255, 300, 350
and 400~Myr are plotted in Fig. 6 along with the CMDs of the cluster.
We adopted $E(B-V)=0.05$  and $(m-M)_{0}=24.64$ while transforming the isochrones 
to the observational plane. 
One can see that none of age/metallicity combinations provides
good fit to the observations. In general, all isochrones are too blue
to reproduce location of cluster upper main sequence in 
F606W/F475W-F606W plane. This may be due to an
intergalactic reddening which was not taken into account.
However, even with assumed $E(B-V)=0.05$ the solar metallicity 
models fail to reproduce location of cluster red giants.
If we neglect the possible intergalactic reddening, 
the most acceptable fit is obtained for ${\rm [Fe/H]}=-0.5$ 
and an age of 300-350 Myr. Assuming a total reddening of 
$E(B-V)\approx 0.12$ one may obtain reasonable fits to observed CMDs
for ${\rm [Fe/H]}=-0.5$  and ${\rm [Fe/H]}=-1.0$. In such a case
the preferred age would be 200-350 Myr.

In Fig. 5 one may see that CMDs of the analyzed $40\times 40$~arcsec$^{2}$ 
field show a pronounced RC. There is no evidence for any
noticeable blue horizontal branch, indicating the paucity of 
old and metal poor stars. Similar morphology
of the CMD was observed by Barker et al. (2008) for 3 fields covering 
south-east outskirts of M33.
The mean luminosity and color of the RC 
is correlated with  average age and metallicity of the stellar
population. The luminosity decreases with increasing  age or 
metallicity,
while the color becomes bluer with decreasing metallicity. 

The deprojected distance of the 33-4-018 field
from the center of M33 is 6.8~kpc.
For RC stars from this field  
we have obtained 
average values of $<M_{I}>=-0.37$ 
and $<(V-I)_{0}>=0.88$. This can be compared with 
results of Barker et al. (2008) 
for 3 other outer fields in M33 observed with  ACS/WFC.
For the field A1 located at deprojected distance  of  9~kpc
they measured $<M_{I}>=-0.41$ and $<(V-I)_{0}>=0.94$,
while for  the outermost field A3 at distance of 13~kpc 
they found $<M_{I}>=-0.38$ and $<(V-I)_{0}>=0.92$.   
If both sets of photometry are indeed on the same photometric
system then one may conclude that old population stars in 
fields A3 and 33-4-018 have similar ages and metallicities. 
This is worth noticing,  given  that the deprojected galactocentric
distances of both field differ by a factor of two.

\Subsection{Cluster SM-57}
The object listed in the SM catalog under number 57 was first noticed
by Christian \& Schommer (1982). They included it on 
a list of unclassified non-stellar objects from M33 region. 
The SM catalog provides just its positional data  along with 
a note that it is a possible galaxy. 
ZKH list the object under the name 33-3-021 and classify it as a 
possible cluster. Examination of ACS/WFC images shows that SM-57 is
indeed a compact stellar cluster. It has an appearance resembling some 
presumed globular clusters from M33 (Sarajedini et al. 2000).
SM-57 has absolute magnitude of $M_{\rm V}\approx -6.4$ ($V\approx 18.4$) 
and half-light radius $r_{h}\approx 3.7$~pc. This places it on 
the $M_{\rm V}/r_{h}$ diagram in the region occupied by several 
Milky Way globular clusters (van den Bergh 2008). However, as we
show below, SM-57 is  a rich intermediate age open 
cluster rather than an old  globular cluster.

The object is visible against the spiral arm III of M33 (Sandage and 
Humphrey 1980). This makes stellar photometry in its field 
difficult, despite a relatively large angular distance of  $d=17.6$~arcmin
from the galaxy center. 
The upper panel of  Fig. 7 shows the CMD for stars located in the 
ring $0.75<r<2.25$~arcsec centered on the 
cluster core. CMD's for  nearby comparison field and 
surrounding  $40\times 40$~arcsec$^{2}$ field are shown in middle and 
bottom panels of Fig. 7, respectively. 
The CMD of the whole field exhibits a well populated
red clump and  an upper red giant branch. One may also notice a relatively
large number of bright main-sequence stars with $F606W<23$ -- 
see Figs. 4 \& 5 for comparison. Apparently, the field hosts a 
noticeable population of young massive stars. 

The available
photometry allows only a very approximate estimation of cluster's age. 
A comparison  of
upper and middle panels of Fig. 7 shows that SM-57 has a well
populated RC. The cluster seems to contain also some bright red giants
with $F606W<24.0$. Such stars are missing from the CMD for the comparison
field.  We note that photometry in the comparison field extends 
about 1 mag deeper relatively to the ring covering the outer part of SM-57. 
This can be explained by increased crowding in the cluster area.
Despite this effect, the cluster CMD shows an excess of stars
at $F606W\approx 26.0$ and $F606W-F814W\approx 0.35$ 
(or $F475W-F606W\approx 0.4$) relatively
to the comparison field. These stars represent likely the top 
of clusters main sequence. In such a case the difference in 
luminosity between the RC and the main sequence top would be
$\Delta F606W\approx \Delta V\approx 1.0$. $\Delta V$ parameter is
a robust  age indicator for intermediate and old stellar clusters
(Cannon 1970). Using calibration of Carraro \& Chiosi (1994)
we obtain for $\Delta V= 1.0$ and assumed ${\rm [Fe/H]}=-0.7$ an 
age of 1.0~Gyr. This estimate provides a rather safe
lower limit on the cluster age. If  the main-sequence of SM-57
were located below $F606W=26.0$, the cluster would be  older than 1~Gyr.
On the other hand, lack of clearly marked giant branch excludes an
age exceeding 4-5 Gyr.

\Subsection{Cluster ZKH 33-6-012}
Like the previous object, ZKH 33-6-012 is visible against the 
spiral arm III of M33. As can be seen in Fig. 2, it 
is located just on the edge of the field observed with ACS/WFC.
The cluster is rather extended, with a total
angular diameter  exceeding 6.2~arcsec which corresponds to a 
linear diameter of 26~pc. In Fig. 8 we show CMD for stars
detected inside radius $r=3.1$~arcsec from the cluster center. 
The middle panel of this figure shows CDMs
for the nearby comparison field having the same area as the 
cluster region. The CMD for $40\times 40$~arcsec$^2$ region
including the cluster is presented in the bottom panel. One may notice
populous RC elongated in the reddening direction. Apparently
this disk field suffers from substantial differential reddening. 
It is interesting that this effect operates even in 
a relatively small region occupied by the  cluster 
(see the upper panel of Fig. 8). 

The cleaned CMD of 33-6-012  is shown in Fig. 9.     
From this plot one may infer that the cluster possess about 60 RC giants. 
This estimate is in fact very conservative. There is no
data for a significant part of the cluster,
  and no correction for incompleteness of 
photometry was applied. With no doubt 33-6-012 is a very 
rich cluster.
Despite this, neither the cluster subgiant branch nor its main-sequence
can be noticed in Fig. 9. Apparently the cluster turn-off is
located below $F606W \approx 26.5$, i.e. more than 1 mag below the RC. 
This implies an age exceeding 1 Gyr. At the same
time lack of noticeable subgiant branch and poorly populated AGB
excludes ages over 4-5 Gyr. We conclude that 33-6-012 is a rich open 
cluster of intermediate age. 

\section{Discussion and conclusions}
By examining archival ACS images of M33 galaxy 
we have confirmed cluster
status for 24 candidates from ZKH catalog. Moreover, 91 new
clusters were detected. Altogether these 115 clusters
expand noticeably a sample of 215 "guaranteed" clusters 
included in the electronic edition of the SM catalog.
The integrated $F606W-F814W$ color was obtained for 108 
objects from our list. Fourt-two of these clusters have
sufficiently red colors to be old globular clusters.
However, one has to bear in mind that there is no unique 
relation between the integrated color and the age of a cluster. 
Figure 4 in Park \& Lee (2007)
shows clearly that on $M_{\rm V}/B-V$ diagram there is a 
substantial overlap of regions occupied by open and globular 
clusters from the Milky Way. 
Hence, red color alone is not sufficient to declare that 
a given cluster from M33 has an age exceeding $\sim 10$ Gyr.

Eight out of 215 "guaranteed" clusters from the SM catalog have
ages estimated at over 10 Gyr. Milky Way harbors about 150 
globular clusters and it is about 6 times more luminous than M33.
One could therefore expect, perhaps a bit naively, that M33 should possess
about 25 old globular clusters.  
Our detection of 10 new candidates for old clusters confirms
the conclusion of SM about a need for more complete and thorough
survey of the whole M33 field. For the moment there are just
18 candidates for globular clusters in M33. 
 
In fact, incompleteness of the sample of M33 clusters is not
the only issue. The question about the actual age of the oldest M33 clusters 
is also far from being answered. Virtually all currently available age 
estimates are based on global parameters of a clusters like integrated
colors and magnitudes (Chandar et al. 1999, 2001; Park \& Lee 2007; SM) 
or integrated spectra (Morettii \& Held 2006). 
Estimates relaying on integral photometry lead to identification
of few clusters with age  exceeding 10 Gyr. Moretti \& Held 
(2006) used Lick indices measured from integrated spectra to derive
age and  metallicity for 42 clusters. They concluded that M33 
globular clusters
are relatively young with ages 8-9 Gyr. So far there are no 
old M33 clusters with resolved stellar photometry 
reaching the main sequence turnoff \footnote{ Two intermediate age open 
clusters with photometry reaching main-sequence turnoff are ZK90 (this paper)
and C38 (Chandar et al. 2006)}.
Sarajedini et al. (2000) obtained V/V-I
diagrams for 10 halo clusters. Their photometry extends 2-2.5 mag
below  the horizontal branch level. This is still 1-1.5 mag above the
level of the main sequence for old globulars.
Moreover, available CMDs 
are heavily contaminated by  field stars. 
For eight out of ten clusters studied by Sarajedini et al. (2000) 
it was possible to state
that horizontal branch is located entirely redward of the 
RR~Lyr instability strip. This points toward  relatively young 
age and/or high metallicity. So far the only M33 cluster known to have a
blue horizontal branch is M9 (Sarajedini et al. 1998; the case
of C20 presented in that paper is in our opinion less convincing).
This cluster has to be as old and metal-poor as some well
studied globular clusters from the Milky Way. 
A presence of old and metal poor population in M33
is further supported by detection of RR~Lyr stars in two outer regions
of the galaxy (Sarajedini et al. 2006). However, actual 
fraction of old globulars and fraction of old stars in 
the M33 disk/halo remains  unknown.  

For two clusters analyzed in some detail in this paper we have 
succeeded with obtaining main-sequence photometry. Cluster ZK-90 turned out 
to have an intermediate age of 3-5 Gyr. Its photometry is 
exceptionally deep thanks to availability of long exposures 
and location in a relatively uncrowded region. 
Cluster 33-4-018, despite being 
located in the outer disk of M33, turned out to be quite young
with an age of 200-350 Gyr. Such an age is not 
surprising, given that young bright and blue stars are 
observed in the disk of M33 at large galactocentric distances
(Barker at al. 2007). Color-magnitude diagrams
of the two remaining clusters show red clump stars
and poorly populated giant branches. This points toward
intermediate ages exceeding 1 Gyr.  

We would like to point out that difficulty 
with obtaining main sequence photometry for old M33
clusters is due not only to  crowding effects. Additional factor
is a marginal sampling of the point spread function provided
by ACS/WFC. This poor sampling makes very difficult the detection
of fainter stars located in wings of brighter stars. 
The ACS/HRC mode offers pixel scale almost two times better
compared to the  WFC mode.  This comes at expense of a slightly reduced 
quantum efficiency.  Possibly, observations in the HRC mode of ACS 
would allow to obtain the first CMDs reaching 
down to the main-sequence for some old M33 clusters.
Among obvious targets for such observation are M33-EC1 (Stonkute et al. 2008)
and 34-5-022 from ZKH. These are two out of three outermost clusters 
known in M33. They are exceptionally extended and uncrowded, and are
indeed ideal candidates for ACS/HRC observations.

\Acknow{We are greatful to Michael Barker for his comments on the
orginal manuscript. Research of JK and KZ is supported by the 
Foundation for the Polish Science through grant MISTRZ. This paper is
based on observations made with the NASA/ESA Hubble  Space Telescope,
and obtained from the Hubble Legacy Archive, which is a collaboration
between the Space Telescope Science Institute (STScI/NASA), the Space
Telescope European Coordinating Facility (ST-ECF/ESA) and the Canadian
Astronomy Data Centre (CADC/NRC/CSA).
}

\begin{figure}[htb]
\centerline{\includegraphics[width=120mm]{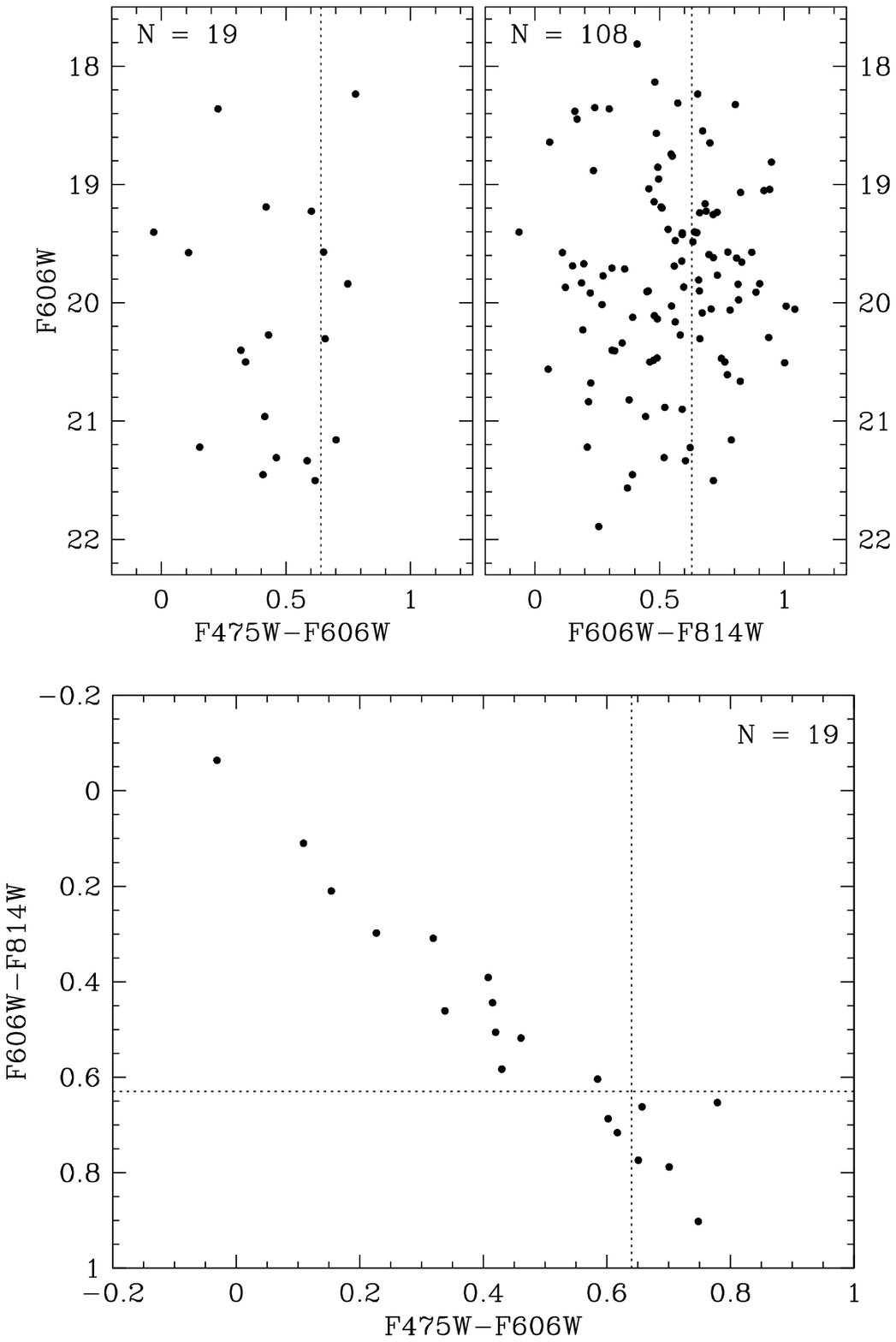}}
\caption{Color-magnitude and color-color diagrams for M33
clusters from Table 2. The dashed lines
mark blue edges of color distribution for globular clusters
from the Milky Way.}
\end{figure}

\begin{figure}[htb]
\centerline{\includegraphics[width=120mm]{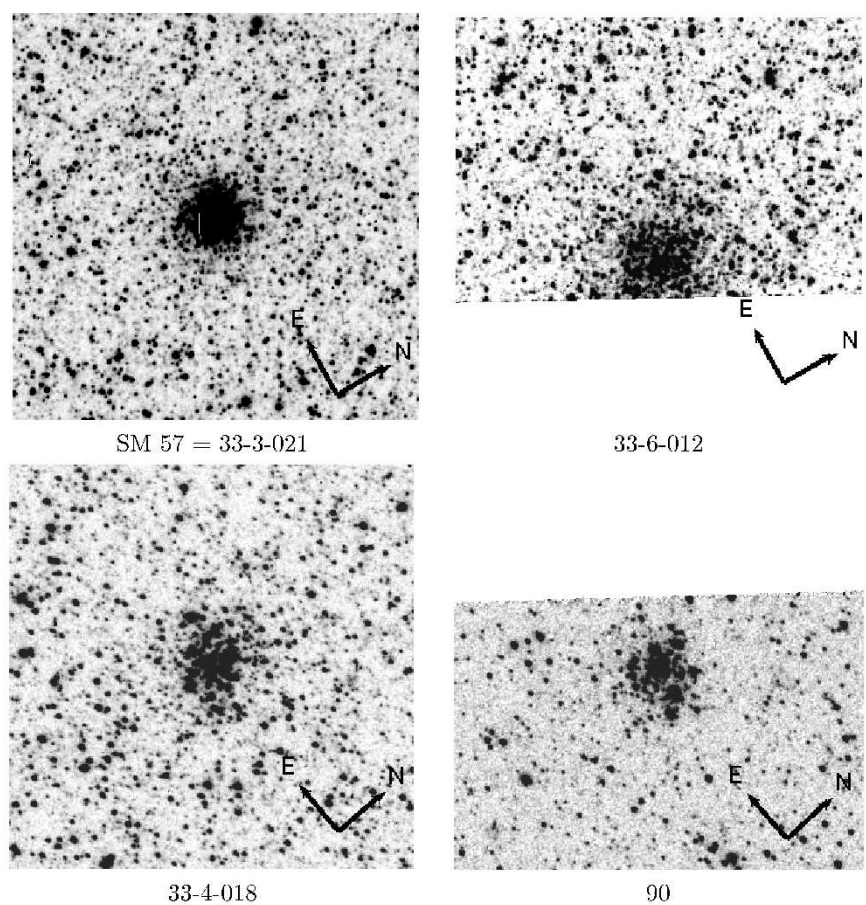}}
\caption{F814W images of clusters analyzed in Sec. 4. All charts are 
16 arcsec wide.}
\end{figure}

\begin{figure}[htb]
\centerline{\includegraphics[width=120mm]{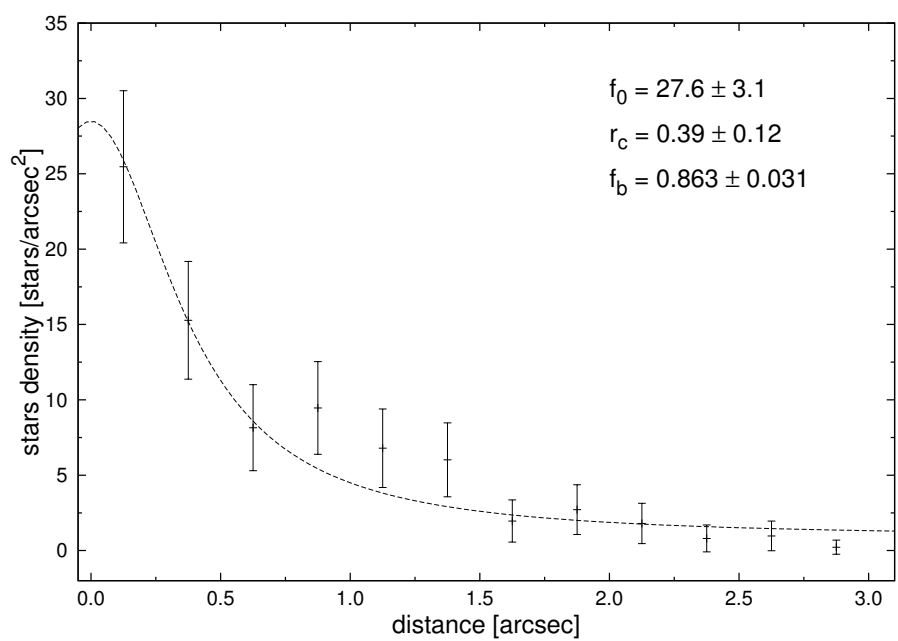}}
\caption{Surface density profile of ZK-90 for stars with
F606W$<$27.5. King's profile fit is shown with the solid line. 
}
\end{figure}

\begin{figure}[htb]
\centerline{\includegraphics[width=120mm]{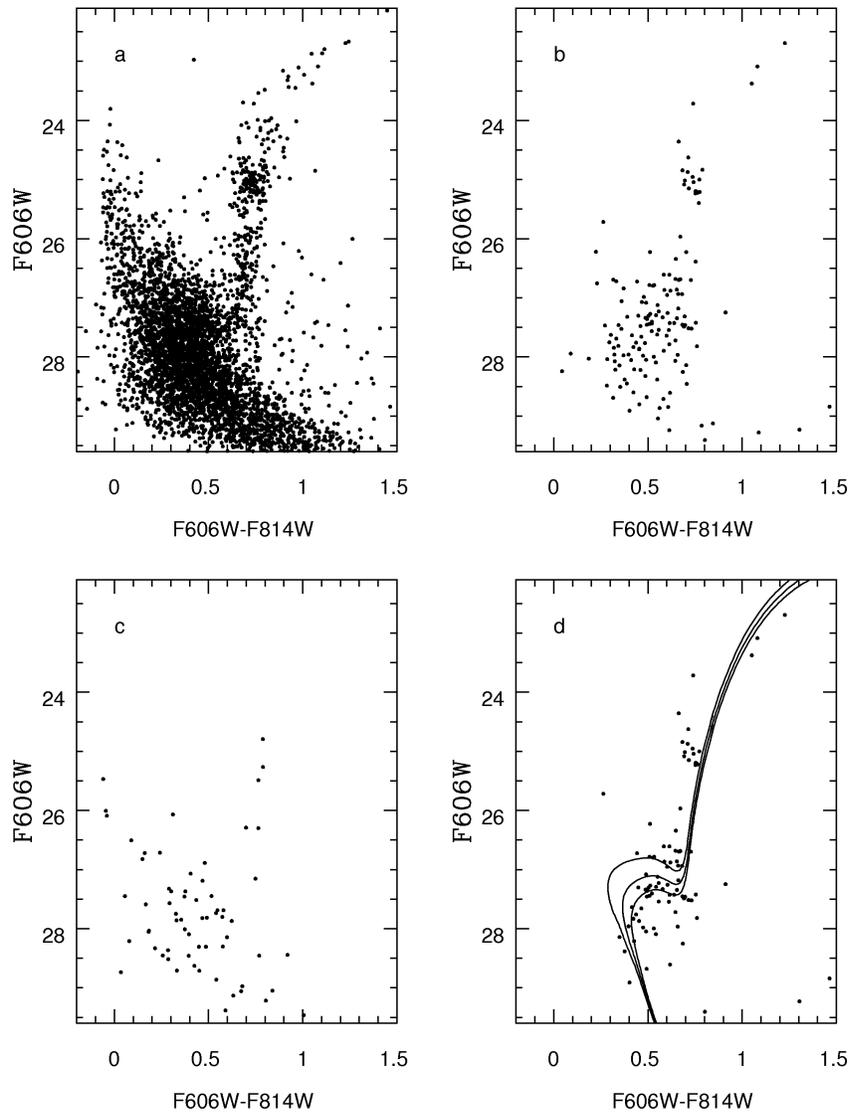}}
\caption{Color magnitude diagrams for the cluster ZK-90
and  the surrounding field. See text for details.
}
\end{figure}

\begin{figure}[htb]
\centerline{\includegraphics[width=120mm]{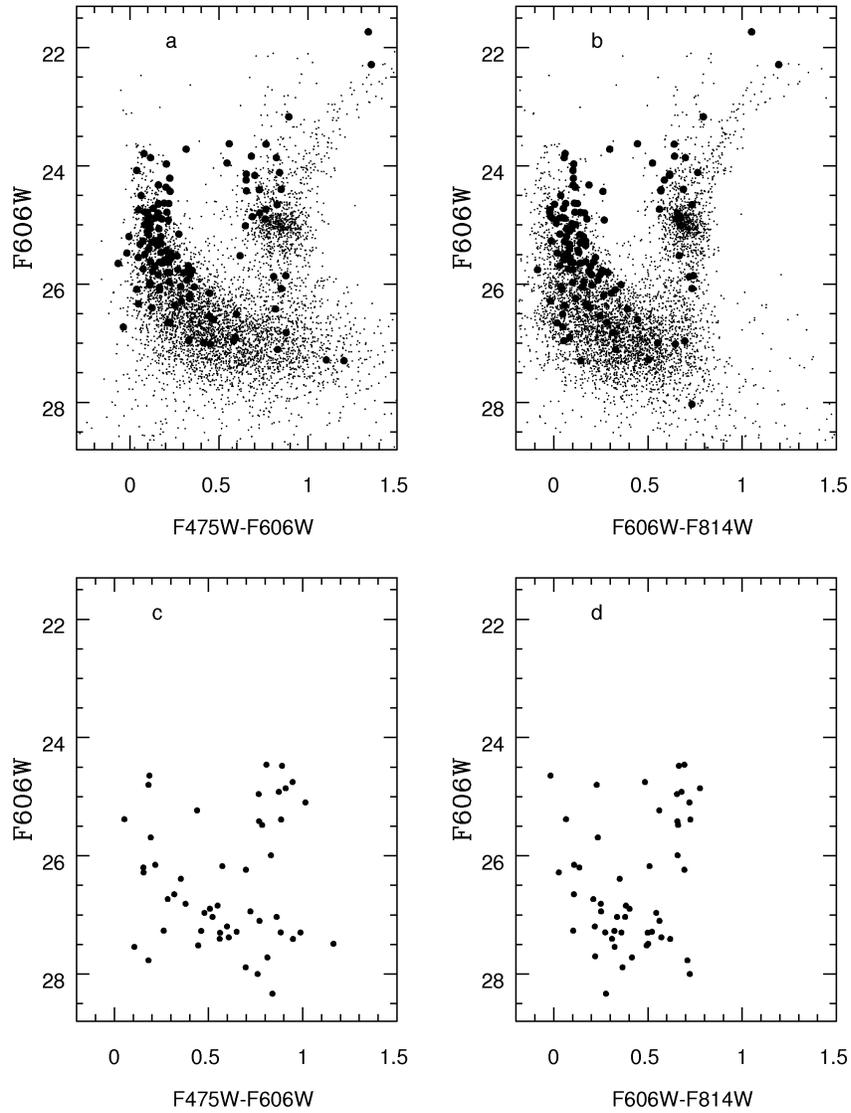}}
\caption{Upper panel: CMDs of the cluster 33-4-018 (filled circles) 
and the surrounding field (small dots).
Lower panel: CMDs for the comparison field having the same area
as the cluster field. 
}
\end{figure}

\begin{figure}[htb]
\centerline{\includegraphics[width=120mm]{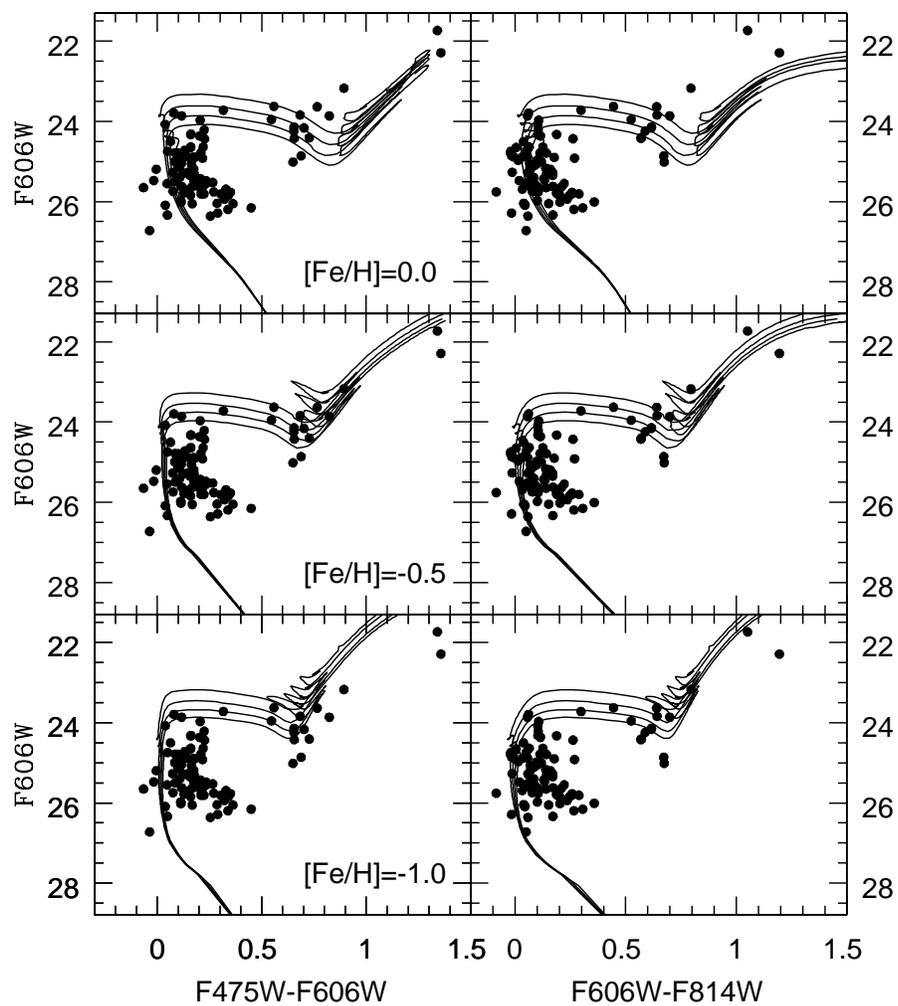}}
\caption{
Field stars corrected CMDs of the cluster 33-4-018. Darthmouth isochrones 
for ages 250, 300, 350 and 400 Myr are marked with continuous lines.
Top, middle and bottom panels show isochrones
for ${\rm [Fe/H]}=0.0$, ${\rm [Fe/H]}=-0.5$ and ${\rm [Fe/H]}=-1.0$,
respectively.  
}
\end{figure}

\begin{figure}[htb] 
\centerline{\includegraphics[width=120mm]{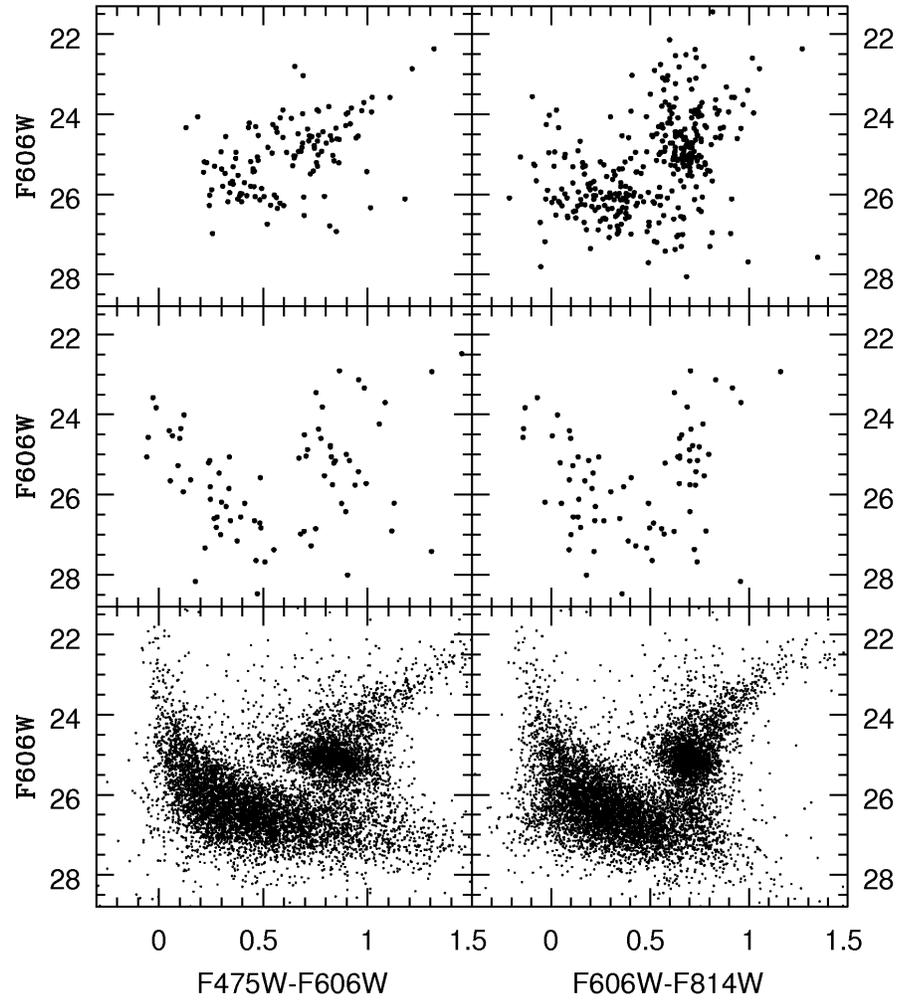}}

\caption{Color-magnitude diagrams for SM-57 (top), the nearby comparison field 
(middle) and $40\times 40$~arcsec$^{2}$ region around the cluster (bottom).
The comparison field has the same area as the cluster field.
}
\end{figure}

\begin{figure}[htb] 
\centerline{\includegraphics[width=120mm]{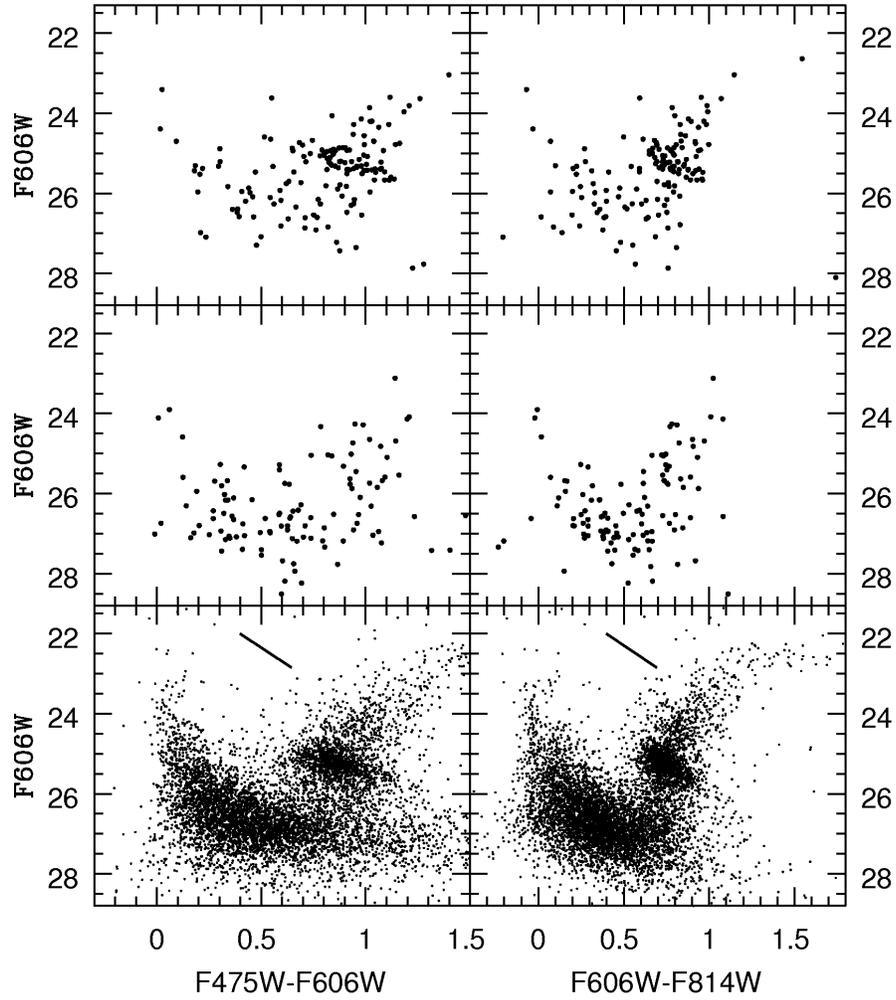}}

\caption{Color-magnitude diagrams for 33-6-012 (top), 
the nearby comparison field 
(middle) and $40\times 40$~arcsec$^{2}$ region around the cluster (bottom).
The reddening vector for $E(B-V)=0.3$ is shown in the lower panel. 
}
\end{figure}

\begin{figure}[htb]

\centerline{\includegraphics[width=120mm]{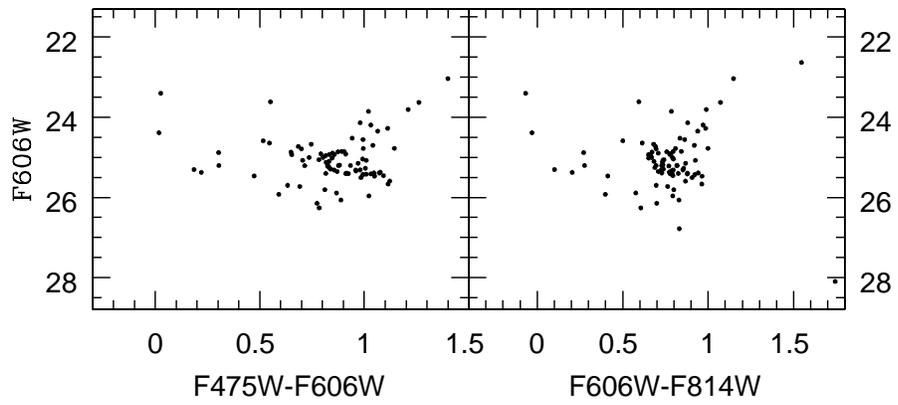}}
\caption{
Cleaned CMDs of the cluster 33-6-012.
}
\end{figure}

\end{document}